\documentclass[%
aip,
preprint,
showpacs,
preprintnumbers,
 amsmath,
 amssymb,
]{revtex4-1}

\usepackage{graphicx}% Include figure files
\usepackage{dcolumn}% Align table columns on decimal point
\usepackage{bm}% bold math
\usepackage{color}
\begin{document}

%\preprint{AIP/123-QED}
\title{Efficient injection of radiation-pressure-accelerated sub-relativistic protons into laser wakefield acceleration based on 10 PW lasers}% Force line breaks with \\
%\thanks{The collisionless shock formation and ion acceleration in intense laser interactions with near-critical-density plasmas}

\author{M. Liu}
\affiliation{Key Laboratory for Laser Plasmas (MoE), School of Physics and Astronomy,
Shanghai Jiao Tong University, Shanghai 200240, China}%
\affiliation{Collaborative Innovation Center of IFSA, Shanghai Jiao Tong University,
Shanghai 200240, China}%

\author{S. M. Weng}\email{wengsuming@gmail.com}%
\affiliation{Key Laboratory for Laser Plasmas (MoE), School of Physics and Astronomy,
Shanghai Jiao Tong University, Shanghai 200240, China}%
\affiliation{Collaborative Innovation Center of IFSA, Shanghai Jiao Tong University,
Shanghai 200240, China}%

\author{H. C. Wang}
\affiliation{Key Laboratory for Laser Plasmas (MoE), School of Physics and Astronomy,
Shanghai Jiao Tong University, Shanghai 200240, China}%
\affiliation{Collaborative Innovation Center of IFSA, Shanghai Jiao Tong University,
Shanghai 200240, China}%

\author{M. Chen}
\affiliation{Key Laboratory for Laser Plasmas (MoE), School of Physics and Astronomy,
Shanghai Jiao Tong University, Shanghai 200240, China}%
\affiliation{Collaborative Innovation Center of IFSA, Shanghai Jiao Tong University,
Shanghai 200240, China}%

\author{Q. Zhao}
\affiliation{Key Laboratory for Laser Plasmas (MoE), School of Physics and Astronomy,
Shanghai Jiao Tong University, Shanghai 200240, China}%
\affiliation{Collaborative Innovation Center of IFSA, Shanghai Jiao Tong University,
Shanghai 200240, China}%

\author{Z. M. Sheng}\email{z.sheng@strath.ac.uk}%
\affiliation{Key Laboratory for Laser Plasmas (MoE), School of Physics and Astronomy,
Shanghai Jiao Tong University, Shanghai 200240, China}%
\affiliation{Collaborative Innovation Center of IFSA, Shanghai Jiao Tong University,
Shanghai 200240, China}%
\affiliation{SUPA, Department of Physics, University of Strathclyde, Glasgow G4 0NG, UK}

\author{M. Q. He}
\affiliation{Institute of Applied Physics and Computational Mathematics, Beijing 100094, China}

\author{Y. T. Li}
\affiliation{Collaborative Innovation Center of IFSA, Shanghai Jiao Tong University,
Shanghai 200240, China}%
\affiliation{National Laboratory for Condensed Matter Physics, Institute of Physics,
Chinese Academy of Sciences, Beijing 100190, China}%
\affiliation{School of Physical Sciences, University of Chinese Academy of Sciences, Beijing 100049, China}

\author{J. Zhang}
\affiliation{Key Laboratory for Laser Plasmas (MoE), School of Physics and Astronomy,
Shanghai Jiao Tong University, Shanghai 200240, China}%
\affiliation{Collaborative Innovation Center of IFSA, Shanghai Jiao Tong University,
Shanghai 200240, China}%

\date{\today}% It is always \today, today,
\begin{abstract}
We propose a hybrid laser-driven ion acceleration scheme using a combination target of a solid foil and a density-tailored background plasma.
In the first stage, a sub-relativistic proton beam can be generated by the radiation pressure acceleration in the intense laser interaction with the solid foil.
In the second stage, this sub-relativistic proton beam is further accelerated by the laser wakefield driven by the same laser pulse in a near-critical-density background plasma with a decreasing density profile.
The propagating velocity of the laser front and the phase velocity of the excited wakefield wave are effectively lowered at the beginning of the second stage.
By decreasing the background plasma density gradually from near critical density along the laser propagation direction, the wake travels faster and faster while it accelerates the protons.
%the phase velocity of the wakefield increases correspondingly.
Consequently, the dephasing between the protons and the wake is postponed, and an efficient wakefield proton acceleration is achieved.
%Because the injection of sub-relativistic protons into the wakefield is ingeniously handled,
This hybrid laser-driven proton acceleration scheme can be realized by using ultrashort laser pulses at the peak power of 10 PW for the generation of multi-GeV proton beams.
% at intensity on the order of $10^{21}$ W/cm$^2$.
\end{abstract}

%\pacs{52.57.Kk, 52.38.Kd, 52.65.Rr}
%52.38.Kd   Laser-plasma acceleration of electrons and ions (see also 41.75.Jv Laser-driven acceleration in electromagnetism; electron and ion optics)
%41.75.Jv   Laser-driven acceleration (see also 52.38.-r Laser-plasma interactions in plasma physics)
%52.50.Jm   Plasma production and heating by laser beams (laser-foil, laser-cluster, etc.)
%52.65.Rr   Particle-in-cell method
%52.27.Ny   Relativistic plasmas
%78.67.Ch 	Nanotubes
%81.07.-b   Nanoscale materials and structures: fabrication and characterization
% PACS, the Physics and Astronomy Classification Scheme.

\maketitle

\section{Introduction}

Plasma-based accelerators driven by ultrashort intense laser pulses have been paid significant interest over the last decades due to their ultrahigh accelerating gradients \cite{Tajima1979,Esarey2009,Daido2012,Macchi2013}.
Aiming at various applications ranging from fundamental sciences to medical and industrial applications, immense progress has been made in accelerating both electrons and ions on the basis of laser-plasma interactions.
%In laser wakefield acceleration (LWFA), electrons can be trapped and continuously accelerated by the plasma wave excited by the ponderomotive force of an ultrashort intense laser pulse.
Recent experiments demonstrate the efficient generation of high-quality multi-GeV electron beams by the laser wakefield acceleration (LWFA) \cite{Wang2013NC,Leemans2014PRL}.
With respect to laser-driven ion acceleration, there are a few of competitive mechanisms depending on laser and target conditions.
In the most experimental conditions, the predominant mechanism is the target normal sheath acceleration (TNSA). In this scenario, a large number of electrons heated by the laser pulse go through the target and then result in a strong charge-separation electrostatic field at the rear of the target \cite{WilksPOP2001}. A recent experiment demonstrates that the ions (preferentially protons) can be accelerated by this electrostatic field to a cut-off energy of about 85 MeV \cite{WagnerPRL2016}.
However, the ion beams generated in the TNSA are usually far from a monoenergetic distribution.
In stark contrast, the generation of a quasi-monoenergetic ion beam could be expected in the scenario of radiation pressure acceleration (RPA) by an ultraintense laser pulse with a typical circular polarization \cite{LS_Esirkepov,MacchiHB,LS_Robinson,LS_Yan}.
In the RPA scenario, the ponderomotive force of an ultraintense laser pulse can make the electrons move forward coherently, which leads to a strong charge-separation field between the preceding electron layer and the laggard ion layer.
The ion layer can be pulled by this charge-separation field and then move together with the electron layer. Depending on the target thickness, the RPA works in two distinct modes.
In the hole-boring mode, a thick solid target can be accelerated layer by layer, and then an ion beam with ultrahigh energy fluence is generated \cite{WengSR}.
In comparison, the acceleration of ions to ultrahigh energies is allowed in the light-sail mode, where a nanoscale solid foil could be continuously accelerated by an ultrashort ultraintense laser pulse under the ideal conditions \cite{LS_Qiao,LS_Chen,LS_Yu}.
However, an obvious gap is disgustingly presented between the experimental results and the theoretical predictions of the RPA. This can be attributed to the onset of transverse instabilities and the restrictions of the laser-target conditions in the experiments \cite{Pegoraro,BinPRL2015,KimPoP2016}.

To obtain a more energetic ion beam, the ion acceleration in the laser-driven wakefield has also been investigated in analogy to the laser wakefield electron acceleration \cite{Shorokhov,Shen2007}. A critical issue of the LWFA is how to keep the particles moving in pace with the accelerating field, i.e., the injection of the particles into the wakefield \cite{ZhaoQian}.
It is typically required that the particles reach the relativistic speed before they are injected into the wakefield.
Since the ions have much higher inertia than the electrons, it becomes more difficult to inject the ions into the wakefield.
The self-injection of protons can only be triggered in the interaction of ultraintense laser pulses with dense plasmas, where the potential difference of the wake is large enough to trap the protons \cite{Shorokhov,Shen2007}.
In a hybrid scheme using a combination target that is composed of a thin solid foil for the RPA and an underdense gas for the LWFA, the injection condition can be greatly relaxed for the relativistic protons that are pre-accelerated by the RPA \cite{LLYu,FLZheng}.
In order to pre-accelerate the protons to relativistic speeds in the RPA stage, however, an ultraintense laser pulse ($\geq 10^{22}$ W/cm$^2$) is usually required, which would crucially postpone the realization of this scheme in experiments.
In addition to the RPA-LWFA hybrid scheme, it was also proposed to enhance the ion energies and/or the ion beam qualities in laser-driven ion acceleration by using some multi-stage schemes \cite{Pfotenhauer,Sinha,Wang,WangPRE2014,Kawata2014,Pei,Kawata2016,Mirzanejhad,WangPOP2017} or phase-matched scheme \cite{Bulanov2010}.

In this paper, we propose a RPA-LWFA hybrid ion acceleration scheme that can work efficiently with an ultrashort moderately intense laser pulse.
With this laser pulse, the protons of a thin solid foil are only pre-accelerated to sub-relativistic speeds by the RPA in the first stage.
In the second stage, the injection of these pre-accelerated sub-relativistic protons into the LWFA is permitted by a near-critical-density plasma whose density decreases gradually from several times to several hundredths of the critical density along the laser propagation direction.
The laser pulse is propagating relatively slow in a classically overdense but relativistically transparent region at the beginning of this stage.
Consequently, the phase velocity of the excited wakefield plasma wave is effectively limited, which is the key to the efficient injection of the sub-relativistic protons into the LWFA.
More importantly, the wakefield wave itself travels faster and faster with the decreasing background plasma density, while accelerating the protons. As a result, the dephasing between the wakefield and the protons is postponed.
With the decreasing background plasma density, the laser depletion is alleviated as well.
Finally, the postponed dephasing and alleviated laser depletion combine to guarantee an efficient LWFA of the protons.

%is analogous to

\section{Theoretical analysis}

\begin{figure}
%\center
\centering
\includegraphics[width=0.40\textwidth, viewport=60 165 495 680, clip]{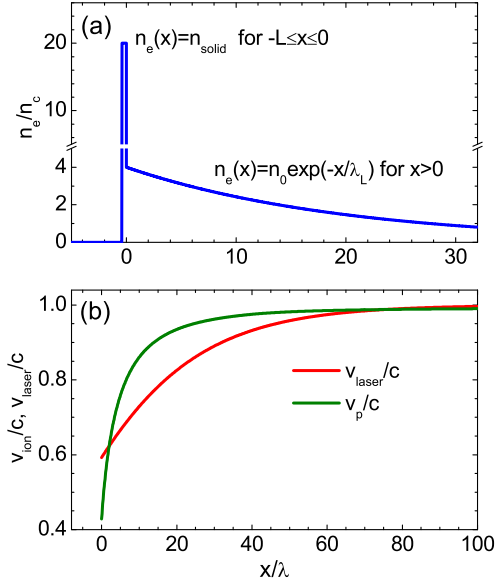}
\caption{(a) The longitudinal density profile of a typical combination target used in the RPA-LWFA hybrid ion acceleration scheme. The target consists of a uniform solid foil with $n_e=n_{solid}=20n_c$ in the $0 \ge x \ge -L= -0.4 \lambda$ region and a near-critical-density background plasma with a tailored density profile $n_e(x)= n_0 \exp(-x/\lambda_L)= 4n_c \exp(-x/20\lambda) $ in the $x>0$ region.
(b) The mean proton velocity $v_p$ and the propagating velocity $v_{laser}$ of the laser front as functions of the coordinate $x$ in the LWFA stage. The mean proton velocity $v_p$ is estimated by Eq. (\ref{Vion}) under the assumptions $a_0=50$ and $v_{p}|_{x=0}\simeq 0.43c$.
The laser front velocity $v_{laser}$ is estimated by Eq. (\ref{Vlaser}) using the tailored density profile of the background plasma in (a).
} \label{figProfile}
\end{figure}

To study the RPA-LWFA hybrid ion acceleration scheme, we begin with some primitive theoretical analysis based on  one-dimensional models. As shown in Fig. \ref{figProfile}, the employed combination target consists of a solid foil and a near-critical-density background plasma with a tailored density profile.
An ultrashort intense circularly polarized (CP) laser pulse is incident from the left side into the combination target.

In the stage of laser-foil interaction, the ion acceleration is dominated by the RPA that can be fairly described by the "light sail" mode  \cite{Macchi2010}.
The obtained mean ion energy, which may be the most interesting parameter for the next acceleration stage, can be estimated as
\begin{equation}
\mathcal{E}_{ion} = m_i c^2 \frac{\Pi ^2}{2(\Pi +1)}, \label{E_ion}
\end{equation}
where $\Pi = 2m_e n_c a_0^2 \tau c / \sum_i m_i n_i L$, the critical density $n_c=m_e\omega^2/4\pi e^2$, the normalized vector potential of the laser pulse $a_0 \equiv|e\textbf{E}/\omega m_e c|$ [for a CP laser pulse $a_0\simeq$($I_0 \lambda^2/2.74\times10^{18}$ W cm$^{-2}$ $\mu$m$^2$)$^{1/2}$], $L$ is the foil thickness,
$\tau$ is the duration of the laser pulse, the sum of $m_in_i$ is over all ion species, and $m_i$ and $n_i$ are the mass and the number density of the i-th ion species, respectively.
In a typical simulation case discussed in the next section, we employ a 33 fs ($\tau=10 T$) laser pulse at the intensity $I_0 \simeq 6.85\times10^{21}$ W cm$^{-2}$ ($a_0=50$), where $T=\lambda/c$ is the laser period and a laser wavelength $\lambda=1$ $\mu$m is assumed. The foil is composed of Carbon and Hydrogen atoms in 1:2 number ratio with $n_e=20 n_c$ and a thickness $L = 0.4\lambda$. Substituting these laser and target parameters into Eq. \ref{E_ion}, one can easily get a mean proton energy $\mathcal{E}_p \simeq 600$ MeV, which corresponds to a sub-relativistic mean proton velocity $v_{p} \simeq 0.79c$.
In reality, the RPA process will be prematurely terminated as long as the target becomes transparent due to the transverse instabilities, the plasma heating and the target deforming.
Therefore, the obtained proton energy is usually much lower than the prediction by Eq. \ref{E_ion}.
For instance, the two-dimensional simulation of the RPA process under the above laser and target parameters only results in a mean proton energy $\mathcal{E}_p \simeq 100$ MeV (the corresponding $v_{p} \simeq 0.43c$).

To inject these sub-relativistic protons into the wakefield, a near-critical-density plasma is applied to limit the phase velocity of the wakefield plasma wave in the second stage.
In the intense laser pulse interaction with this dense plasma, a large amplitude wakefield can be excited with a structure of channel-like electron bubble \cite{Shen2007}.
Thanks to its large amplitude, the LWFA will boost the speed of the sub-relativistic protons dramatically.
From the energy conservation, one can get
\begin{equation}
\frac{d \mathcal{E}_{p} }{d x} = m_p c^2 \frac{d \gamma_{p} }{d x} = e E_x(x), \label{energyCon}
\end{equation}
where $\gamma_{p}=1/(1-v_{p}^2/c^2)^{1/2}$ is the Lorentz factor of the protons, and $E_x(x)$ is the local accelerating field felt by the protons.
For simplicity, we assume that $E_x(x)\sim E_{max}/2$, where $E_{max} \approx a_0 \omega m_e c/e $ is a rough approximation for the amplitude of the longitudinal electric field in the relativistic transparency regime \cite{Shorokhov}.
Under these assumption and approximation, Eq. (\ref{energyCon}) can be rewritten as
\begin{equation}
\frac{d \gamma_{p} }{d x} \approx  \frac{m_e}{m_p} \frac{\pi a_0 }{ \lambda}.  \label{Vion}
\end{equation}
Substituting $v_{p,0} \simeq 0.43c$ ($\mathcal{E}_{p} \simeq 100$ MeV) obtained in the RPA stage as the initial proton velocity in the LWFA and $a_0=50$ into the above equation, a rapid rise in the proton velocity $v_{p}$ at the early stage of the LWFA is evidenced in Fig. \ref{figProfile}(b).
This is different from the LWFA of relativistic protons \cite{LLYu,FLZheng}, whose velocities have been initially close to the speed of light in the vacuum and thus cannot increase obviously.

For the LWFA of sub-relativistic protons, nevertheless, the phase velocity of the wakefield wave must be simultaneously boosted in order to avoid an early dephasing between the wakefield and the protons.
In a near-critical-density plasma, the wake phase velocity $v_{wake}$ is approximate to the propagating velocity $v_{laser}$ of the laser front.
The latter can be estimated as
\begin{equation}
v_{laser} \sim \exp(-\frac{4n_e}{n_{cr}})(1-\frac{n_e}{n_{cr}})^{1/2} c , \label{Vlaser}
\end{equation}
where $n_{cr}\simeq (1+0.48a^2_0)^{1/2}n_c$ is the relativistic critical density for CP pulses at intensities $a_0 \gg 1$ \cite{Weng2012,WengPoP2012}.
The above expression indicates that $v_{laser}$ increases with a decreasing plasma density, so does the wake phase velocity $v_{wake}$.
For simplicity, we assume an exponentially decreasing plasma density $n_e(x)=n_0 \exp(-x/\lambda_L)$, where $n_0$ and $\lambda_L$ are selected according to the initial proton velocity and the derivative of the proton velocity over $x$.
As shown in Fig. \ref{figProfile}(a), we set $n_0=4n_c$ and $\lambda_L=20 \lambda$ in a typical simulation case.
Fig. \ref{figProfile}(b) indicates that in this case the laser front velocity $v_{laser}$ can increase gradually as the mean proton velocity.

\begin{figure}
%\raggedleft
\centering
  \includegraphics[width=0.4\textwidth, viewport=55 155 500 690, clip]{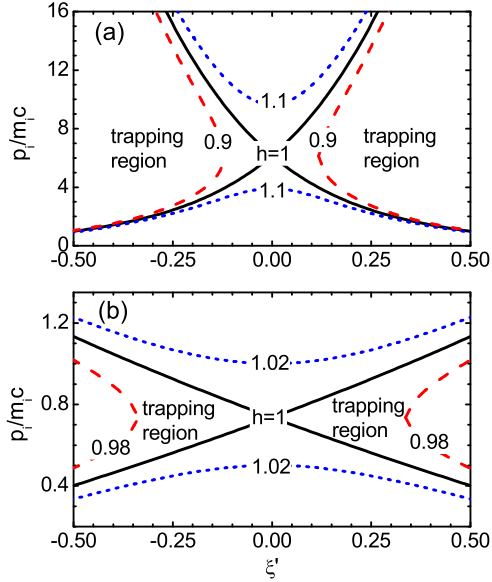}
\caption{
The trapping region of the ions in the $\xi' - p'_i$ phase space at two different plasma densities (a) $n_e=0.1n_c$, and (b) $n_e=4n_c$. The boundary of the trapping region is defined by $h(\xi',p'_i)=1$ (black solid lines), and an ion will be trapped by the wake if its Hamiltonian $h(\xi',p'_i)<1$ (such as red dashed lines). The laser amplitude is assumed to be $a_0=50$.
} \label{figTrapping}
\end{figure}

To illuminate the necessity of a near-critical-density plasma for the trapping of sub-relativistic protons, the trapping region in the proton phase space at different plasma densities are compared in Fig. \ref{figTrapping}.
In the co-moving frame ($\xi=x-v_{wake}t$) with the wake, the ion motion is governed by a conservative Hamiltonian \cite{Shorokhov}
\begin{equation}
h(\xi',p'_i)=\gamma_{w} \left[(1+p'^2_i)^{1/2}-p'_i \frac{v_{wake}}{c} - \frac{m_e n_c a_0^2}{2m_in_e} \xi'^2 \right], \nonumber \label{Hamilton}
\end{equation}
where $\xi'=\xi \omega_p/c$ and $p'_i=p_i/m_ic$ are the normalized coordinate and momentum of the ion, the Lorentz factor $\gamma_{w}=1/(1-v_{wake}^2/c^2)^{1/2}$ is corresponding to the wake phase velocity $v_{wake}$, and $v_{wake}\simeq v_{laser}$.
The equation $h(\xi',p'_i)=1$ defines the boundary of the trapping region in the $\xi' - p'_i$ phase space. An ion will be trapped by the wake if its Hamiltonian $h(\xi',p'_i)<1$.
In an underdense plasma with $n_e=0.1n_c$, the trapping condition $h(\xi',p'_i)<1$ is only satisfied for relativistic ions with $p_i/m_ic \geq 1$ as shown in Fig. \ref{figTrapping}(a).
In contrast, Fig. \ref{figTrapping}(b) illustrates that a sub-relativistic ion with $p_i \approx 0.4 m_i c$ is already possible to enter into the trapping region in a near-critical-density plasma with $n_e=4n_c$.
In such a near-critical-density plasma, both the larger amplitude and the lower phase velocity of the wakefield play a positive role in the trapping of sub-relativistic ions \cite{Shorokhov,Shen2007}.

\section{Simulation results}

\begin{figure}
%\raggedleft
\centering
  \includegraphics[width=0.48\textwidth, viewport=90 175 500 670, clip]{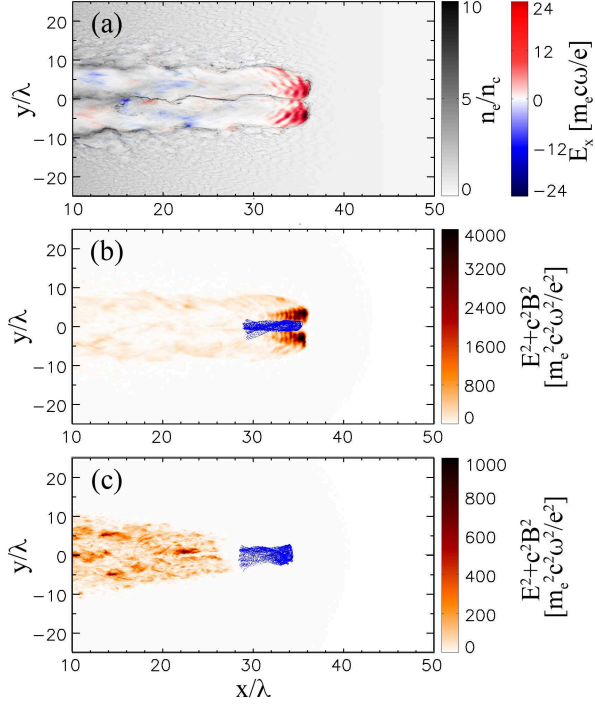}
\caption{
(a) Distributions of the background electron density $n_e$ (gray) and the longitudinal electric field $E_x$ (blue-red), and (b) the energetic protons with $\mathcal{E}_{p}\ge 500$ MeV (blue dots) and the laser energy distribution (red temperature) at $t=50T$ obtained in the simulation using the combination target of a solid foil and a density-tailored background plasma.
(c) The energetic protons  with $\mathcal{E}_{p}\ge 500$ MeV (blue dots) and the laser energy distribution at $t=50T$ obtained in the simulation using the combination target of a solid foil and a uniform background plasma.
The foil has an electron density $n_e=20 n_c$ and a thickness $L = 0.4\lambda$. The background plasma has either a uniform density profile $n_e=4 n_c$ or a tailored density profile $n_e=4 n_c \exp(-x/20\lambda)$.
The laser pulse has an amplitude $a_0=50$, a duration $\tau=10T$, and a waist $\sigma=8 \lambda$.
} \label{figPhysics}
\end{figure}

To verify the efficiency of the RPA-LWFA hybrid ion acceleration using a tailored plasma density profile, we have carried out a series of two-dimensional particle-in-cell (PIC) simulations using the code OSIRIS \cite{Fonseca}.
The moving-window technique is employed to reduce the computational loads.
In each simulation, a $70 \times 70 $ $\mu$m$^2$ simulation box moves along the $x$-axis at the speed of light, and it is divided into $5600 \times 4200$ cells.
The combination target is composed of a thin solid foil in the front region ($-0.4\lambda \le x \le 0$ and $|y| \le 2.5 \lambda$) and a semi-infinite background plasma at $x \ge 0$.
The foil is composed of Carbon and Hydrogen atoms in 1:2 number ratio with $n_e=n_{proton}+6n_{carbon}=20 n_c$.
Here, we assume the foil to have a relatively low density but a large thickness in order to lower computational cost.
The foil with a higher density but a smaller thickness should reproduce similar results, if the areal density of the foils remains the same \cite{LLYu}.
The semi-infinite background plasma only contains Carbon atoms, and the comparison is made between the background plasmas with a uniform density profile $n_e=4 n_c$ and a tailored density profile $n_e=4 n_c \exp(-x/20\lambda)$.
Each cell has 400 macro-particles per cell in the foil region, and 100 macro-particles per cell in the semi-infinite background plasma region.
The laser pulse is incident from the left side into the combination target, and it is circularly polarized and has a wavelength $\lambda=1$ $\rm{\mu m}$.
Its intensity profile is assumed to be $a(t,y) = a_0 \sin^2(\pi t/ 2\tau ) \exp(-y^2/\sigma^2)$ for $0\le t \le 2\tau $. In typical simulations, we set the amplitude $a_0=50$, the duration $\tau=10T$, and the waist $\sigma=8 \lambda$.
For reference, the front of the laser pulse is assumed to arrive at the vacuum-foil interface ($x=-0.4\lambda$) at t=0.

Figure \ref{figPhysics}(a) illustrates that a bubble of electron density has been formed in the laser interaction with the background plasma. In this near-critical-density plasma, however, the electron bubble is not closed at the tail while shows a channel-like structure \cite{Shen2007}. Accompanying the channel-like bubble formation, a longitudinal electric field whose amplitude is on the order of the laser amplitude is generated \cite{Shorokhov}.
In addition, this longitudinal electric field is highly asymmetrical, which is in contrast to the quasi-symmetric wakefields in tenuous plasmas.
In a near-critical-density plasma, the wake is characterized by a strong positive longitudinal electric field at the front part, while the negative longitudinal electric field at the rear part is relatively weak.
Such an asymmetrical structure of the longitudinal electric field is of great benefit to accelerating the ions.
As shown in Fig. \ref{figPhysics}(b), a large number of protons have been accelerated by this wakefield to energies higher than 500 MeV at $t=50T$.
More importantly, the laser pulse still travels together with these energetic protons since the front velocity of the former is boosted in the background plasma with a decreasing density.
Consequently, these energetic protons can be further accelerated by the wakefield stimulated by the laser pulse.
In a uniform near-critical-density plasma, however, the LWFA of the protons has already ceased at $t=50T$ since these energetic protons overrun the laser pulse as shown in Fig. \ref{figPhysics}(c).
Fig. \ref{figPhysics}(c) also illustrates that the laser depletion is much more severe in this uniform near-critical-density plasma with $n_e=4n_c$ \cite{WengPoP2012}.
While the laser depletion is greatly alleviated in the background plasma with a decreasing density as shown in Fig. \ref{figPhysics}(b), which is also crucially for the efficient LWFA of the ions.

\begin{figure}
%\raggedleft
\centering
 \includegraphics[width=0.4\textwidth, viewport=45 160 490 690, clip]{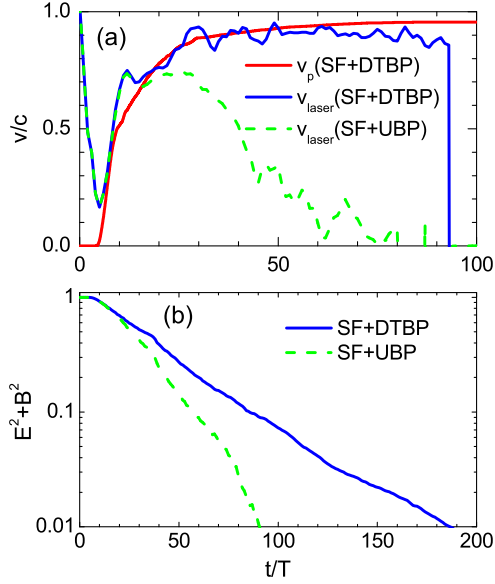}
\caption{(a)Temporal evolution of the propagating velocities of the laser fronts ($v_{laser}$) obtained in the simulations using the combination targets of a solid foil and a uniform background plasma (SF+UBP), or of a solid foil and a density-tailored background plasma (SF+DTBP), respectively. (b) Temporal evolution of the total electromagnetic field energy ($E^2+B^2$) in these two cases, where the total energy is normalized to the initial total energy of the incident laser pulse. The velocity ($v_p$) of an accelerated proton in the case with a density-tailored background plasma is also drawn in (a) for comparison. The simulation parameters are the same as those in Fig. \ref{figPhysics}.
} \label{figVelocities}
\end{figure}

The temporal evolutions of the laser front velocities in the laser interactions with different combination targets are compared in Fig. \ref{figVelocities}(a). The propagating velocity of the laser front is defined as the forward velocity of the longitudinal coordinate $x_f$ where the laser intensity $I(x_f)=I_0/100$, here $I_0$ is the initial laser intensity.
In the first stage ($t \le 12T$), the propagating velocity of the laser front is independent on the background plasma since the laser pulse mainly interacts with the thin solid foil. In this stage the propagating velocity of the laser front at first decreases dramatically after the laser pulse hits the solid foil and then increases quickly as long as the foil is accelerated by the RPA.
In the second stage ($t \ge 12T$), it is clear that the propagating velocity of the laser front substantially depends on the density profile of the background plasma since the thin foil becomes transparent.
In the uniform background plasma with $n_e=4n_c$, the propagating velocity of the laser front at first remains quasi-steady and then decreases gradually due to the laser depletion.
In contrast, in the background plasma with a decreasing density the propagating velocity of the laser front increases gradually until the laser pulse is depleted.
For comparison, the time evolution of the velocity of a typical proton accelerated in this case is also displayed in Fig. \ref{figVelocities}(a).
It is evident that the propagating velocity of the laser front can increase nearly at the same rate as the velocity of accelerated protons for a quite long time in this case.
As a result, the dephasing between the accelerated protons and the wakefield is effectively postponed in the background plasma with an exponentially decreasing density profile.
Further, the temporal evolution of the total energy of the laser pulse with these two different combination targets are compared in Fig. \ref{figVelocities}(b). It is confirmed that the laser depletion can be greatly alleviated in the background plasma with an exponentially decreasing density profile.

\begin{figure}
%\raggedleft
\centering
 \includegraphics[width=0.4\textwidth, viewport=40 140 490 680, clip]{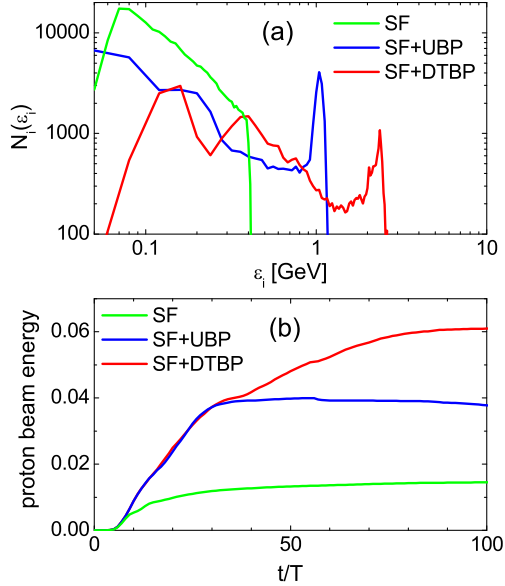}
\caption{(a) The proton energy spectra obtained with three different target configurations: the single solid foil (SF), the combination target of a solid foil and a uniform background plasma (SF+UBP), and the combination target of a solid foil and a density-tailored background plasma (SF+DTBP).
(b) The time evolutions of the corresponding proton beam energies in these three cases, where the proton beam energy is normalized to the total energy of the incident laser pulse.
The simulation parameters are the same as those in Fig. \ref{figPhysics}.
} \label{figEnergy}
\end{figure}

Thanks to the postponed dephasing and the alleviated laser depletion, the protons can be efficiently accelerated up to higher energies by the LWFA in the background plasma with an exponentially decreasing density profile.
The proton energy spectra obtained with different target configurations are compared in  Fig. \ref{figEnergy}(a), while the time evolutions of the total energies of the proton beams are displayed in Fig. \ref{figEnergy}(b).
In the laser interaction with a single thin foil, a large amount of protons can be accelerated by the RPA. However, the RPA ceases as long as the thin foil becomes transparent.
So the peak energy of the final proton beam is only about 100 MeV, and the corresponding mean proton velocity is sub-relativistic ($0.43c$). In this case, the energy conversion efficiency from the laser pulse to the proton beam is relatively low ($\sim1.5\%$).
By attaching a uniform background plasma with $n_e=4n_c$ to the rear of the thin foil, some protons that are pre-accelerated by the RPA can be trapped and further accelerated by the LWFA in the attached background plasma. Consequently, a quasi-monoenergetic proton beam with a peak energy about 1 GeV is generated. Meanwhile, the energy conversion efficiency from the laser pulse to the proton beam increases to about $4\%$ in this hybrid acceleration scheme.
However, the time evolution of the proton beam energy in Fig. \ref{figEnergy}(b) implies that the LWFA of the protons quickly ends at $t=30T$ due to the early dephasing between the protons and the wakefield wave in this case.
If a tailored density profile $n_e(x)=4n_c\exp(-x/20\lambda)$ is employed for the background plasma, the LWFA of the protons in the background plasma can persist much longer and become more efficient. As a result, the peak energy of the final quasi-monoenergetic proton beam can be doubled to 2 GeV. Correspondingly, the final energy conversion efficiency from the laser pulse to the proton beam is boosted up to $6\%$.

\section{Discussion and Conclusion}

In the above simulations, the laser pulse is assumed to be ultrashort (33 fs) and moderately intense ($I_0 \simeq 6.85\times 10^{21}$ W/cm$^2$). Such a laser pulse is well within the capacity of kJ 10 PW laser facilities, such as ELI, OMEGA EP, and Apollon \cite{Papadopoulos}.
Using such a moderately intense laser pulse, however, it is difficult to accelerate the protons directly to a highly relativistic speed by the RPA.
In this case, a near-critical-density background plasma is critically required for the injection of the sub-relativistic protons accelerated by the RPA into the LWFA.
In a near-critical-density plasma, the wake phase velocity is approximate to the propagating velocity of the laser front, while the latter is substantially determined by the energy balance rather than the conventional dispersion relation \cite{WengPoP2012}.
With a decreasing density profile, the propagating velocity of the laser front is allowed to increase gradually in a near-critical-density background plasma, so does the phase velocity of the excited wake.
As a result, the wake can travel with the accelerated protons for a long time in the RPA-LWFA hybrid ion acceleration scheme using the combination target with a density-tailored background plasma.

\begin{figure}
%\raggedleft
\centering
 \includegraphics[width=0.4\textwidth, viewport=40 140 490 680, clip]{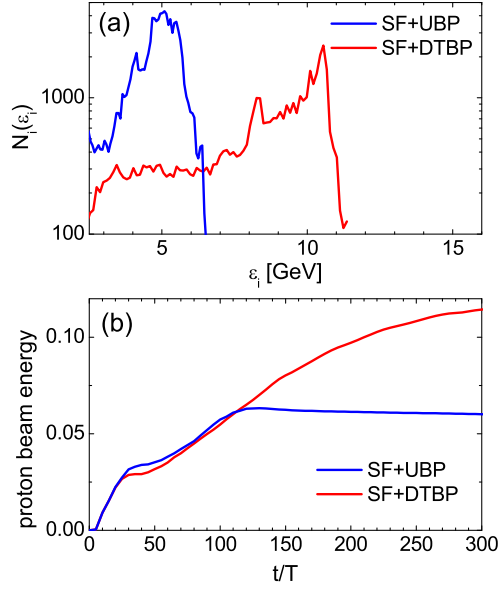}
\caption{(a) The proton energy spectra obtained with two different combination targets: a solid foil and a uniform background plasma (SF+UBP), or a solid foil and a density-tailored background plasma (SF+DTBP).
(b) The time evolutions of the corresponding proton beam energies in these two cases, where the proton beam energy is normalized to the total energy of the incident laser pulse.
The foil has an electron density $n_e=20 n_c$ and a thickness $L = 1 \lambda$. The background plasma has either a uniform density profile $n_e=1.2 n_c$ or a tailored density profile $n_e=1.2 n_c \exp(-x/60\lambda)$.
The laser pulse has an amplitude $a_0=120$, a duration $\tau=10T$, and a waist $\sigma=8 \lambda$.
} \label{figUltraintense}
\end{figure}

Although this ion acceleration scheme is primarily designed for the moderately intense lasers, it also possesses an apparent advantage at ultrahigh laser intensities that may be available in the near future.
In the simulations presented in Fig. \ref{figUltraintense}, an ultrahigh laser intensity $I_0\simeq 4\times 10^{22}$ W/cm$^2$ ($a_0=120$) is employed while other laser parameters remain the same as in the previous simulations.
Correspondingly, the thickness of the solid foil is modified to $L = 1 \lambda$ in order to optimize the RPA in the first stage.
Using such an ultraintense laser pulse, the protons can be pre-accelerated by the RPA to higher velocities.
So the background plasma density should be correspondingly reduced to result in a higher propagating velocity of the laser pulse in the LWFA stage.
Meanwhile, the gradient of the background plasma density should also be reduced since the proton velocities have already been close to the speed of the light in the vacuum.
Consequently, we find that the efficient LWFA can be achieved in the background plasma with a modified density profile $n_e=1.2 n_c \exp(-x/60\lambda)$ in this case.
The resultant proton energy spectrum and the time evolutions of the proton beam energy are drawn in Fig. \ref{figUltraintense}, where the results using the combination target with a uniform background plasma ($n_e=1.2 n_c$) are also displayed for comparison.
It is illustrated that in the RPA-LWFA hybrid ion acceleration scheme the mean energy of the final proton beam can be boosted from about 5 GeV to more than 10 GeV if the uniform background plasma is replaced by the density-tailored background plasma. In the meantime, the energy conversion efficiency from the laser pulse to the final proton beam increases from $\sim 6.0\%$ to $\sim 11.4\%$.

%Since the final proton energy from LWFA depends strongly on the rate of laser depletion, more detail on the latter process should be given, e.g., by showing the evolution of the total energy and/or the peak intensity of the laser pulse. For the same reason, the dependence of the laser depletion time on the pulse duration should be given and discussed in the context of the present acceleration scheme.

\begin{figure}
\centering
\includegraphics[width=0.40\textwidth,viewport=65 410 490 690, clip]{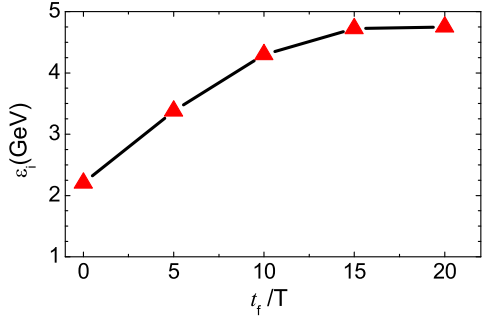}
\caption{The peak energy of the final proton beam as a function of the flat-top part length ($t_f$) of a temporally tailored laser pulse. The tailored laser pulse has a trapezoidal ($10T$ rise + $t_f$ flat-top + $10T$ fall) temporal intensity profile.
The rising edge, the falling edge and other parameters of the laser pulse are the same as those in Fig. \ref{figPhysics}.
The target is same to the combination target of a solid foil and a density-tailored background plasma used in Fig. \ref{figPhysics}.} \label{difflaser}
\end{figure}

As the laser depletion time is roughly proportional to the pulse duration, it seems that the final proton energies should be further enhanced by using a longer laser pulse. However, we find that a longer pulse duration has two mutually offsetting effects. On the one hand, it may prolong the LWFA stage due to the lengthened laser depletion time. On the other hand, it will reduce the efficiency of the RPA since the laser pondermotive force of a longer pulse is relatively weaker.
Therefore, the final proton energies nearly don't increase with the increasing pulse duration if a Gaussian laser pulse is still assumed.
Nevertheless, a longer laser pulse could be of great benefit if its temporal intensity profile can be tailored to have a sharp rising edge.
For instance, a trapezoidal ($10T$ rise + $t_f$ flat-top + $10T$ fall) temporal intensity profile can be assumed by inserting a flat-top part into the previous Gaussian laser pulse.
Fig. \ref{difflaser} illustrates that the peak energy of the final proton beam can be greatly enhanced up to $\sim 4.7$ GeV by increasing the duration of such a temporally tailored laser pulse.
The peak energy of the proton beam will be saturated because the dephasing finally will terminate the LWFA even if the laser pulse is not depleted.

It is worth pointing out that the near-critical-density plasma behind the foil not only provides the background plasma for the LWFA, but may also stabilize the RPA via replenishing the electrons to the foil \cite{WengSR,Shen2017NJP}.
Accidentally, the electron injection into the LWFA can also be enhanced with a decreasing plasma density profile \cite{Gonsalves}. However, the underlying principle is different. For the LWFA of electrons in a tenuous plasma, the phase velocity of the wake is transiently reduced by decreasing the plasma density, which allows the electrons to catch up and then be trapped in the wake.

%The Authors are recommended to add some comments how to produce the density tailored plasma behind the solid foil. In addition, additional brief discussions on the sensitivity of the density tailoring on the ion acceleration would be welcome.

In experiments, the density-tailored plasma behind the solid foil might be achieved with the help of a secondary nanosecond laser pulse at a lower intensity \cite{Humieres}.
Irradiated by this longer, lower-intensity laser pulse, the thin foil will be exploded.
And the density profile of the exploded foil can be tuned by varying the energy of this longer laser pulse and the time delay between this longer pulse and the main pulse \cite{Humieres}.
Alternatively, such a density-tailored near-critical-density plasma might be obtained by ionization of an ultra low-density plastic foam if its density can be modulated \cite{ChenSN}.

\begin{figure}
\centering
\includegraphics[width=0.40\textwidth,viewport=40 410 490 680, clip]{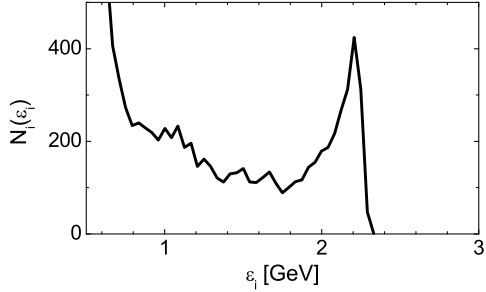}
\caption{
The proton energy spectrum obtained with an alternative combination target of a solid foil and a density-tailored background plasma with a semi-Gaussian density profile $n_e(x)=3 n_c \exp [-x^2/(30\lambda)^2]$. This spectrum is similar to the one in the case when the background plasma has an exponentially decreasing density profile. The parameters of the laser pulse and the solid foil are the same as those in Fig. \ref{figPhysics}.} \label{semiGaussian}
\end{figure}

In addition, it is worth mentioning that the efficiency of this RPA-LWFA hybrid ion acceleration should be not very sensitive to the density fluctuation of the density-tailored background plasma. This is because the efficiency of this hybrid ion acceleration is mainly determined by the dephasing and the laser depletion, both of which are the integral effects over the distance. Therefore, the local density fluctuation nearly doesn't affect the final acceleration efficiency as long as the scale length of the overall density profile is in the appropriate range.
Furthermore, we find that this hybrid ion acceleration could also work very efficiently for the density-tailored background plasmas with some other kinds of density profiles instead of an exponentially decreasing density profile. For instance, Fig. \ref{semiGaussian} illustrates that the background plasma with a semi-Gaussian density profile $n_e(x)=3 n_c \exp [-x^2/(30\lambda)^2]$ could also result in a quasi-monoenergetic spectrum similar to that in the case when the background plasma has an exponentially decreasing density profile.

To summarize, in the intense laser interaction with a near-critical-density plasma a large amplitude wakefield wave can be stimulated, whose phase velocity is effectively lowered. This leads to an efficient injection of sub-relativistic protons, and a rapid rise in the speeds of these sub-relativistic protons.
By decreasing the plasma density gradually, the phase velocity of the wakefield wave can increase simultaneously with the proton speeds, which postpones the dephasing between the protons and the wake. As a result, the protons can be further accelerated by the LWFA for a long time.
Based on this, we propose a RPA-LWFA hybrid proton acceleration scheme using a combination target of a solid foil and a near-critical-density plasma with a decreasing density profile.
The protons of the foil are only needed to be accelerated to sub-relativistic speeds by the RPA in the first stage.
In the second stage, these sub-relativistic protons are injected and further accelerated by the LWFA in a near-critical-density plasma with a decreasing density profile.
Benefited from the efficient injection of sub-relativistic protons into the LWFA, this hybrid acceleration scheme works efficiently even with ultrashort laser pulses at intensity on the order of $10^{21}$ W/cm$^2$.

\begin{acknowledgments}
The work was supported by the National Basic Research Program of China (Grant No. 2013CBA01504), National Natural Science Foundation of China (Grant Nos. 11675108, 11655002, 11721091, 11535001, and 11774227), National 1000 Youth Talent Project of China, Science and Technology Commission of Shanghai Municipality (Grant No. 16DZ2260200).
Simulations have been carried out on the Pi supercomputer at Shanghai Jiao Tong University.
\end{acknowledgments}

\bibliography{apssamp}% Produces the bibliography via BibTeX.

\begin{thebibliography}{9}

\bibitem{Tajima1979} T. Tajima and J. M. Dawson, Phys. Rev. Lett. \textbf{43} 267 (1979).

\bibitem{Esarey2009} E. Esarey, C. B. Schroeder and W. P. Leemans, Rev. Mod. Phys. \textbf{81}, 1229 (2009).

\bibitem{Daido2012} H. Daido, M. Nishiuchi and A. S Pirozhkov, Rep. Prog. Phys. \textbf{75}, 056401 (2012).

\bibitem{Macchi2013} A. Macchi, M. Borghesi and M. Passoni, Rev. Mod. Phys. \textbf{85}, 751 (2013).

\bibitem{Wang2013NC} X. M. Wang, R. Zgadzaj, N. Fazel, Z. Y. Li, S. A. Yi, X. Zhang, W. Henderson, Y. Y. Chang, R. Korzekwa, H. E. Tsai, C. H. Pai, H. Quevedo, G. Dyer, E. Gaul, M. Martinez, A. C. Bernstein, T. Borger, M. Spinks, M. Donovan, V. Khudik, G. Shvets, T. Ditmire and M. C. Downer, Nat. Comms. \textbf{4}, 1988 (2013).

\bibitem{Leemans2014PRL} W. P. Leemans, A. J. Gonsalves, H. S. Mao, K. Nakamura, C. Benedetti, C. B. Schroeder, Cs. T\'{o}th, J. Daniels, D. E. Mittelberger, S. S. Bulanov, J. L. Vay, C. G. R. Geddes and E. Esarey, Phys. Rev. Lett. \textbf{113}, 245002 (2014).

\bibitem{WilksPOP2001} S. C. Wilks, A. B. Langdon, T. E. Cowan, M. Roth, M. Singh, S. Hatchett, M. H. Key, D. Pennington, A. MacKinnon and R. A. Snavely, Phys. Plasmas. \textbf{8}, 542 (2001).

\bibitem{WagnerPRL2016} F. Wagner, O. Deppert, C. Brabetz, P. Fiala, A. Kleinschmidt, P. Poth, V. A. Schanz, A. Tebartz, B. Zielbauer, M. Roth, T. St\"{o}hlker and V. Bagnoud, Phys. Rev. Lett. \textbf{116}, 205002 (2016).

\bibitem{LS_Esirkepov} T. Esirkepov, M. Borghesi, S. V. Bulanov, G. Mourou and T. Tajima, Phys. Rev. Lett. \textbf{92}, 175003 (2004).

\bibitem{MacchiHB} A. Macchi, F. Cattani, T. V. Liseykina and F. Cornolti, Phys. Rev. Lett. \textbf{94}, 165003 (2005).

\bibitem{LS_Robinson} A. P. L. Robinson, M. Zepf, S. Kar, R. G. Evans and C. Bellei, New J. Phys. \textbf{10}, 013021 (2008).

\bibitem{LS_Yan} X. Q. Yan, C. Lin, Z. M. Sheng, Z. Y. Guo, B. C. Liu, Y. R. Lu, J. X. Fang and J. E. Chen, Phys. Rev. Lett. \textbf{100}, 135003 (2008).

\bibitem{WengSR} S. M. Weng, M. Liu, Z. M. Sheng, M. Murakami, M. Chen, L. L. Yu and J. Zhang, Sci Rep. \textbf{6}, 22150 (2016).

\bibitem{LS_Qiao} B. Qiao, M. Zepf, M. Borghesi and M. Geissler, Phys. Rev. Lett. \textbf{102}, 145002 (2009).

\bibitem{LS_Chen} M. Chen, A. pukhov and T. P. Yu, Phys. Rev. Lett. \textbf{103}, 024801 (2009).

\bibitem{LS_Yu} T. P. Yu, A. Pukhov, G. Shvets and M. Chen, Phys. Rev. Lett. \textbf{105}, 065002 (2010).

\bibitem{Pegoraro} F. Pegoraro and S. V. Bulanov, Eur. Phys. J. D. \textbf{55}, 399-405 (2009).

\bibitem{BinPRL2015} J. H. Bin, W. J. Ma, H. Y. Wang, M. J. V. Streeter, C. Kreuzer, D. Kiefer, M. Yeung, S. Cousens, P. S. Foster, B. Dromey, X. Q. Yan, R. Ramis, J. Meyer-ter-Vehn, M. Zepf and J. Schreiber, Phys. Rev. Lett. \textbf{115}, 064801 (2015).

\bibitem{KimPoP2016} I. J. Kim, K. H. Pae, II. W. Choi, C. L. Lee, H. T. Kim, H. Singhal, J. H. Sung, S. K. Lee, H. W. Lee, P. V. Nickles, T. M. Jeong, C. M. Kim and C. H. Nam, Phys. Plasmas \textbf{23}, 070701 (2016).

\bibitem{Shorokhov} O. Shorokhov and A. Pukhov, Laser and Particle Beams, \textbf{22}, 175-181 (2004).

\bibitem{Shen2007} B. F. Shen, Y. L. Li, M. Y. Yu and J. Cary, Phys. Rev. E \textbf{76}, 055402 (2007).

\bibitem{ZhaoQian} Q. Zhao, S. M. Weng, Z. M. Sheng, M. Chen, G. B. Zhang, W. B. Mori, B. Hidding, D. A. Jaroszynski and J. Zhang, New J. Phys., \textbf{20}, 063031 (2018).

\bibitem{LLYu} L. L. Yu, H. Xu, W. M. Wang, Z. M. Sheng, B. F. Shen, W. Yu and J. Zhang, New J. Phys.,\textbf{12}, 045021 (2010).

\bibitem{FLZheng} F. L. Zheng, H. Y. Wang, X. Q. Yan, T. Tajima, M. Y. Yu and X. T. He, Phys. Plasmas \textbf{19}, 023111 (2012).

\bibitem{Pfotenhauer} S. M. Pfotenhauer, O. J\''{a}ckel, J. Polz, S. Steinke, H. P. Schlenvoigt, J. Heymann, A. P. L. Robinson and M. C. Kaluza, New J. Phys. \textbf{12}, 103009 (2010).

\bibitem{Sinha} U. Sinha, Phys. Plasmas \textbf{20}, 073116 (2013).

\bibitem{Wang} W. P. Wang, B. F. Shen, X. M. Zhang, X. F. Wang, J. C. Xu, X. Y. Zhao, Y. H. Yu, L. Q. Yi, Y. Shi, L. G. Zhang, T. J. Xu, and Z. Z. Xu, Phys. Plasmas \textbf{20}, 113107 (2013).
%\bibitem{Wang} W. P. Wang, B. F. Shen, and X. M. Zhang \emph{et al.}, Phys. Plasmas \textbf{20}, 113107 (2013).

\bibitem{WangPRE2014} H. Y. Wang, C. Lin, B. Liu, Z. M. Sheng, H. Y. Lu, W. J. Ma, J. H. Bin, J. Schreiber, X. T. He, J. E. Chen, M. Zepf and X. Q. Yan, Phys. Rev. E, \textbf{89}, 013107 (2014).

\bibitem{Kawata2014} S. Kawata, D. Sato, T. Izumiyama, T. Nagashima, M. Takano, D. Barada, Y. Y. Ma, W. M. Wang, Q. Kong, P. X. Wang and Y. J. Gu, JPS Conf. Proc. \textbf{1}, 015089 (2014).

\bibitem{Pei} Z. K. Pei, B. F. Shen, X. M. Zhang, W. P. Wang, L. G. Zhang, L. Q. Yi, Y. Shi, and Z. Z. Xu, Phys. Plasmas \textbf{22}, 073113 (2015).
%\bibitem{Pei} Z. K. Pei, B. F. Shen, and X. M. Zhang \emph{et al.}, Phys. Plasmas \textbf{22}, 073113 (2015).

\bibitem{Kawata2016} S. Kawata, D. Kamiyama, Y. Ohtake \emph{et al.}, J. Phys.: Conf. Ser. \textbf{691}, 012021(2016).

\bibitem{Mirzanejhad} S. Mirzanejhad, A. Joulaei and J. Babaei, Phys. Plasmas \textbf{23}, 123108 (2016).

\bibitem{WangPOP2017} H. C. Wang, S. M. Weng, M. Murakami, Z. M. Sheng, M. Chen, Q. Zhao and J. Zhang, Phys. Plasmas \textbf{24}, 093117 (2017).

\bibitem{Bulanov2010} S. V. Bulanov, E. Yu. Echkina, T. Zh. Esirkepov, I. N. Inovenkov, M. Kando, F. Pegoraro, and G. Korn, Phys. Rev. Lett. \textbf{104}, 135003 (2010).


\bibitem{Macchi2010} A. Macchi, S. Veghini, T. V Liseykina and F. Pegoraro, New J. Phys. \textbf{12}, 045013 (2010).

\bibitem{Weng2012} S. M. Weng, M. Murakami, P. Mulser and Z. M. Sheng, New J. Phys. \textbf{14}, 063026 (2012).

\bibitem{WengPoP2012} S. M. Weng, P. Mulser and Z. M. Sheng, Phys. Plasmas \textbf{19}, 022705 (2012).

\bibitem{Fonseca} R. A. Fonseca, L. O. Silva, F. S. Tsung, V. K. Decyk, W. Lu, C. Ren, W. B. Mori, S. Deng, S. Lee, T. Katsouleas and J. C. Adam, Lect. Not. Comput. Sci. \textbf{2331}, 342 (2002).

\bibitem{Papadopoulos}  D. N. Papadopoulos, J. P. Zou, C. Le Blanc, G. Ch\'{e}riaux, P. Georges, F. Druon, G. Mennerat, P. Ramirez, L. Martin, A. Fr\'{e}neaux, A. Beluze, N. Lebas, P. Monot, F. Mathieu, and P. Audebert, High Power Laser Sci. Eng. \textbf{4}, e34 (2016).

\bibitem{Shen2017NJP} X. F. Shen, B. Qiao, H. X. Chang, W. L. Zhang, H. Zhang, C. T. Zhou and X. T. He, New J. Phys. \textbf{19}, 033034  (2017).

\bibitem{Gonsalves} A. J. Gonsalves, K. Nakamura, C. Lin, D. Panasenko, S. Shiraishi, T. Sokollik, C. Benedetti, C. B. Schroeder, C. G. R. Geddes, J. van Tilborg, J. Osterhoff, E. Esarey, C. Toth and W. P. Leemans, Nature Physics \textbf{7}, 862 (2011).

\bibitem{Humieres} E. {d}'Humi\`{e}res, P. Antici, M. Glesser, J. Boeker, F. Cardelli, S. Chen, J. L. Feugeas, F. Filippi, M. Gauthier, A. Levy, P. Nicola\"{\i}, H. P\'{e}pin, L. Romagnani, M. Scisci\`{o}, V. T. Tikhonchuk, O. Willi, J. C. Kieffer and J. Fuchs, Plasma Phys. Control. Fusion \textbf{55}, 124025 (2013).

\bibitem{ChenSN} S. N. Chen, T. Iwawaki, K. Morita, P. Antici, S. D. Baton, F. Filippi, H. Habara, M. Nakatsutsumi, P. Nicola\"{\i}, W. Nazarov, C. Rousseaux, M. Starodubstev, K. A. Tanaka and J. Fuchs, Sci. Reports \textbf{6}, 21495 (2016).

%\bibitem{MLiu} M. Liu, S. M. weng, Y. T. Li, D. W. Yuan, M. Chen, P. Mulser, Z. M. Sheng, M. Murakami, L. L. Yu, X. L. Zheng and J. Zhang,Phys. Plasmas \textbf{23}, 113103 (2016).

\end{thebibliography}

\end{document}